ORIGINAL RESEARCH ARTICLE

# Development and Comparison of Model-Based and Data-Driven Approaches for the Prediction of the Mechanical Properties of Lattice Structures

*Chiara Pasini, Oscar Ramponi, Stefano Pandini, Luciana Sartore, and Giulia Scalet*

Lattice structures have great potential for several application fields ranging from medical and tissue engineering to aeronautical one. Their development is further speeded up by the continuing advances in additive manufacturing technologies that allow to overcome issues typical of standard processes and to propose tailored designs. However, the design of lattice structures is still challenging since their properties are considerably affected by numerous factors. The present paper aims to propose, discuss, and compare various modeling approaches to describe, understand, and predict the correlations between the mechanical properties and the void volume fraction of different types of lattice structures fabricated by fused deposition modeling 3D printing. Particularly, four approaches are proposed: (i) a simplified analytical model; (ii) a semi-empirical model combining analytical equations with experimental correction factors; (iii) an artificial neural network trained on experimental data; (iv) numerical simulations by finite element analyses. The comparison among the various approaches, and with experimental data, allows to identify the performances, advantages, and disadvantages of each approach, thus giving important guidelines for choosing the right design methodology based on the needs and available data.

**Keywords** additive manufacturing, lattice structures, neural networks

## 1. Introduction

Lattice structures are three-dimensional architectures made of one or more repeating unit cells (Ref 1). Thanks to their architecture, these structures have mechanical, thermal, and functional properties unachievable by their base materials, such as lightweight, high strength, and enhanced energy absorption, that make them suitable candidates for several application fields from medical and tissue engineering to automotive, mechanical, aeronautical, and seismic ones (Ref 2-4). Moreover, the continuing advancements on additive manufacturing (AM) technologies allow to overcome manufacturing issues typical of standard processes and to produce tailored designs with the highest freedom (Ref 5, 6).

The properties of lattice structures are considerably affected by numerous factors, including unit cell topology and relative density, boundary and loading conditions, presence of defects, and fabrication process parameters (Ref 7). Consequently, designing lattice structures with tailored properties is challenging due to the numerous variables involved.

This challenge motivates efforts to propose reliable methodologies for the design, prediction, and optimization of the behavior and performances of lattice structures, thus avoiding the high costs typical of fabrication and experimentation.

Among the available methodologies, both model-based and data-driven approaches have been proposed (Ref 8, 9). Finite element models of either the entire lattices or single unit cells (Ref 10-28) as well as homogenization approaches, studying the macroscopic behavior of the heterogenous lattice material by an "equivalent" homogenous one (Ref 29, 30), represent the most used tools for accurately predicting and simulating the behavior of these structures. An adjoint variable method has been proposed in (Ref 31) in order to improve the computational efficiency for molecular dynamics with many design variables. Alternative approaches for reducing computational expense, but often at the cost of accuracy, e.g., beam-based or Gibson-Ashby models, have been proposed (Ref 32-43). Topology and shape optimization approaches have been employed to design lattice structures satisfying application requirements (Ref 29, 44-47). Machine learning methods do not need to develop analytical equations and have potential as a time-efficient substitute for finite element analysis (FEA), but they learn the relation between the input or design parameters and the output performance based on the existing data. They **Chiara Pasini, Stefano Pandini,** and **Luciana Sartore**, Department of Mechanical and Industrial Engineering, University of Brescia, via Branze 38, 25133 Brescia, Italy; and **Oscar Ramponi** and **Giulia Scalet**, Department of Civil Engineering and Architecture, University of Pavia, via Ferrata 3, 27100 Pavia, Italy. Contact e-mail: giulia.scalet@unipv.it.

have been employed both for predicting the mechanical characteristics of additively manufactured microstructures (Ref 48, 49), also in the presence of defects (Ref 50), and for the inverse design of new topologies featuring specific mechanical properties (Ref 49, 51-56). Recently, Hussain et al. (Ref 57) have proposed a data-driven constitutive model based on an artificial neural network (ANN) to represent the complex mechanical behavior of a new three-dimensional lattice-structured material.

It is clear that it becomes important to choose the right methodology when approaching the design of lattice structures.

Motivated by the examined framework, the present paper aims to propose, discuss, and compare various modeling approaches to describe and predict the correlations between the mechanical properties and the geometrical features of different types of three-dimensional lattice structures fabricated by fused deposition modeling (FDM).

To this purpose, four approaches are proposed and analyzed: (i) a simplified analytical model; (ii) a semi-empirical model combining analytical equations with experimental correction factors; (iii) an ANN trained on experimental data; (iv) numerical simulations by finite element analyses. The approaches here proposed represent different alternatives to the Gibson-Ashby power model which was found in previous literature on lattice structures (Ref 1, 5, 19, 32-43, 58) and consisted in providing a correlation of their relative modulus and relative strength with their relative density by fitting experimental data. The reason behind the choice of focusing on these approaches is twofold: first, the paper aims to analyze recent data-driven approaches and their application to lattice structures; second, it aims to provide a comparison with more traditional ones, discussing both advantages and disadvantages.

Lattice structures from (Ref 58) are here complemented with additional combinations of their geometrical parameters, and the experimental data on their mechanical properties are correlated with their void volume fraction and used for the calibration of data-driven approaches (ii) and (iii) and the validation of all four.

In particular, the four approaches are employed not only to predict the correlations for the experimentally-tested structures, but also for untested ones. The comparison of the different strategies will be also helpful in guiding the designer in choosing the right methodology based on the needs and available data.

The paper is organized as follows. Section 2 presents the investigated lattice structures, while Sect. 3 discusses the experimental methods. Section 4 presents the adopted modeling approaches. Results are presented and discussed in Sect. 5. Finally, conclusions and perspectives are given in Sect. 6.

## 2. Investigated Lattice Structures

The investigated strut-based lattice structures are shown in Fig. 1. Each structure is composed of a periodic arrangement of a unit cell along the x, y, and z axes. In particular, cubic, tetragonal, body-centered cubic (with internal diagonals), and face-centered cubic (with wall diagonals) unit cells are analyzed. Accordingly, labels SC, ST, BCC, and FCC are used to denote the resulting structures. Strut-based lattices have attracted increasing interest due to their easy design and manufacture, as well as their high structural performance (Ref 19). Cubic and tetragonal unit cells are among the most common and simple in design, and the choice of these specific four types aims at considering the presence of differently oriented struts as well as different unit cell aspect ratios. Their specific strut arrangements also allow to develop predictive models with different degrees of complexity and to more easily understand the lattice mechanical behavior, also when compared with the predicted response.

Table 1 summarizes the main geometric features of each investigated structure. As it can be observed, the wall thickness, $t$, is kept fixed and equal to 0.6 mm, while the hole width, $w$, and the hole height, $h$, are variable parameters (Fig. 1). For SC, BCC, and FCC lattices, $h$ is equal to $w$. Accordingly, the structures are labeled by means of the following notation: (i) SC_w for cubic unit cells; (ii) ST_w_h for tetragonal unit cells; (iii) BCC_w for body-centered cubic unit cells (with internal diagonals); (iv) FCC_w for face-centered cubic unit cells (with wall diagonals). The cells are then repeated periodically along the x, y, and z axes (z corresponding to the build direction). The repetition is defined by the number of holes along these axes ($N$ along x and y; $M$ along z). The repetition is chosen in such a way to generate a lattice structure having, as much as possible, an external cubic volume equal to 1000 mm$^3$ (side lengths of 10 mm). For each specimen, the theoretical void volume fraction, $V_{v,th}$, is evaluated as:

$$V_{v,\mathrm{th}}[\%] = \left(1 - \frac{V_{s,\mathrm{th}}}{V_{\mathrm{th}}}\right) \times 100 \qquad (\mathrm{Eq}\ 1)$$

where $V_{s,\mathrm{th}}$ is the volume occupied by the struts, computed by the CAD model, and $V_{\mathrm{th}}$ is the external cubic volume of the structure (i.e., $A \times A \times H$, according to Fig. 1).

The last column of Table 1 ("Exp. Test") specifies whether the structures have been subjected to mechanical tests (see details in Sect. 3.3) or not; the number and the direction (either xy or z) refer to the number of specimens tested along that direction (see details in Sect. 3.3). Untested configurations are introduced as unseen inputs for the ANN presented in Sect. 4.3.

## 3. Experimental Methods

### 3.1 3D Printing

The structures described in Sect. 2 are realized using fused deposition modeling (FDM) 3D printing, following the protocol described in (Ref 58). The specimens are first drawn using Solidworks software (Dassault Systèmes, Vélizy-Villacoublay, France) and the STL files are elaborated through IdeaMaker slicer software (Raise 3D Technologies, Inc., Irvine, CA, United States). Then, the Raise3D Pro2 printer (build volume of 305 × 305 × 300 mm), equipped with a nozzle of 0.2 mm in diameter, is employed. The printed material is a white RAISE3D Premium PLA filament of 1.75 mm in diameter. The layer thickness is 0.1 mm. The flow rate percentage is 100%, with a printing speed of 25 mm/s. The nozzle temperature is equal to 205 °C and the bed temperature to 60 °C. These process parameters were selected based on the supplier's specifications (Ref 59) and on the small size of the lattice struts, which oriented the choice toward a thin nozzle diameter and a low layer height, looking for better printing precision. The build direction is parallel to the z axis (Fig. 1) for all the specimens realized.

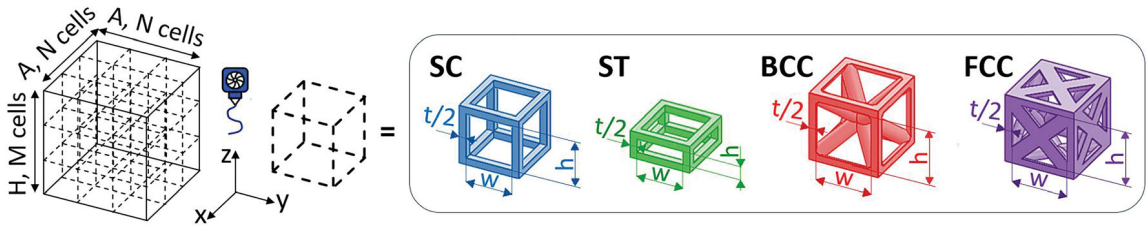

**Fig. 1** Designs of the investigated lattice structures

## 3.2 Experimental Void Volume Fraction

For each specimen, the experimental value of the void volume fraction, $V_{v,\exp}$, is also evaluated as:

$$V_{v,\exp}[\%] = \left(1 - \frac{V_{s,\exp}}{V_{\exp}}\right) \times 100 \quad \text{(Eq 2)}$$

where $V_{s,\exp}$ is the volume occupied by PLA struts and $V_{\exp}$ is the external cubic volume of the printed structure. In particular, $V_{s,\exp}$ is calculated as $m/\rho_0$, where $m$ is the mass of the specimen, measured by laboratory balance (Gibertini E42-B, Gibertini Elettronica, Novate Milanese, Italy), and $\rho_0$ is the declared density of Raise3D PLA Premium [i.e., 1.2 g/cm$^3$ (Ref 59)]; $V_{\exp}$ is obtained from caliper measurements of the lattice outer dimensions in the three directions x, y, and z.

## 3.3 Mechanical Characterization

Compression tests are carried out on the 3D printed structures by means of an electromechanical dynamometer (Instron, Mod. 3366, Illinois Tool Works Inc., Norwood, MA, United States) with a 10 kN load cell. The specimens are compressed at room temperature between flat and parallel plates, with a crosshead speed of 2 mm/min, until a steeply rising portion appears in their load vs. displacement curves, due to densification of the cellular structure. Two types of tests are conducted, one by applying the load along the transversal direction, x or y (denoted, in the following, as xy), and one by applying the load along the build direction, z (see Fig. 1). A total of 64 tests are thus conducted (repetitions are indicated in Table 1).

Figure 2 schematically represents two of the most typical trends of the apparent stress, $\sigma_{app}$, vs. apparent strain, $\varepsilon_{app}$, curves, on which the compressive apparent modulus, $E_{app}$, and compressive apparent failure stress, $\sigma_{app,f}$, are evaluated. Particularly, the apparent stress, $\sigma_{app}$, is evaluated as a nominal value by dividing the measured force, $F$, by the outer initial cross section of the specimen. The outer initial cross section is computed as $A \times A$ for samples tested in z direction and as $A \times H$ for samples tested in xy direction (Fig. 1). The apparent strain, $\varepsilon_{app}$, is evaluated as the ratio between the crosshead displacement and the initial height of the specimen. The initial height is assumed equal to $H$ for samples tested in z direction and as $A$ for samples tested in xy direction (Fig. 1). As schematically represented in Fig. 2, the compressive apparent modulus, $E_{app}$, is defined as the initial slope of the $\sigma_{app}$ vs. $\varepsilon_{app}$ curve, while the compressive apparent failure stress, $\sigma_{app,f}$, as the stress related to the first evidence of sample failure (related to buckling phenomena and/or yielding). Such evidence is displayed as a marked deviation from linearity, forming a "knee" or a peak in the curve. Since the considered loading directions are two, the measured quantities are denoted as $E_{app,xy}$ and $\sigma_{app,f,xy}$ for the x (or y) testing direction and $E_{app,z}$ and $\sigma_{app,f,z}$ for the z testing direction.

## 4. Modeling Approaches

This section presents the four approaches used to describe the correlations between the mechanical properties and the void volume fraction of the lattice structures reported in Table 1.

### 4.1 Analytical Model

The first approach is an analytical model that recalls the rule of mixtures (ROM) typically employed in composite materials (Ref 60). Only lattice struts parallel to the loading direction are considered as actually contributing to the material stiffness and strength; according to the analogy with composite materials, these struts may be seen as uniaxial composite fibers inside a matrix with zero stiffness and strength. By applying the ROM to the modulus ($E_{ROM}$) and to the failure stress ($\sigma_{ROM,f}$), the following equations are obtained:

$$E_{ROM}[\text{MPa}] = E_c V_{//} \quad \text{(Eq 3)}$$

$$\sigma_{ROM,f}[\text{MPa}] = \sigma_{c,f} V_{//} \quad \text{(Eq 4)}$$

where $E_c = 2000$ MPa and $\sigma_{c,f} = 59$ MPa are, respectively, the modulus and the failure stress of the bulk material (i.e., 3D-printed PLA), while $V_{//}$ is the volume fraction of struts parallel to the loading direction. Particularly, 3D-printed PLA modulus, $E_c$, is identified based on FEA of structure SC_3 (see Sect. 4.4), while 3D-printed PLA compressive strength, $\sigma_{c,f}$, is calculated as 125% of the tensile strength in the datasheet of Raise3D Premium PLA (Ref 59). $V_{//}$ can be also written as the ratio between the area occupied by the cross sections of struts parallel to the loading direction and the total cross section of the specimen.

**4.1.1 Analytical Equations for SC and ST Unit Cells.** Considering SC and ST unit cells, the ratio between the mechanical properties of the lattices and those of PLA results as follows:

$$\frac{E_{ROM}}{E_c} = \frac{\sigma_{ROM,f}}{\sigma_{c,f}} = V_{//} = \begin{cases} \frac{t^2}{(w+t)^2} & \text{along } z \\ \frac{t^2}{(w+t)(h+t)} & \text{along } xy \end{cases} \quad \text{(Eq 5)}$$

The relative density of SC and ST unit cells can be calculated as the ratio between the volume occupied by their struts and the total cell volume, resulting in the following equation:

**Table 1** Geometrical dimensions of the investigated structures, according to definitions given in Fig. 1

| Structure label | Diagonal | t, mm | w, mm | h, mm | N | M | $V_{th}$, mm³ | $V_{v,th}$, % | Exp. test |
|---|---|---|---|---|---|---|---|---|---|
| SC_0.5 | None | 0.6 | 0.5 | 0.5 | 9 | 9 | 1158 | 39 | xy (3) z (3) |
| SC_0.7 | None | 0.6 | 0.7 | 0.7 | 8 | 8 | 1331 | 51 | xy (1) z (1) |
| SC_1 | None | 0.6 | 1 | 1 | 6 | 6 | 1061 | 63 | xy (3) z (3) |
| SC_1.5 | None | 0.6 | 1.5 | 1.5 | 5 | 5 | 1368 | 75 | xy (1) z (1) |
| SC_2 | None | 0.6 | 2 | 2 | 4 | 4 | 1331 | 82 | xy (3) z (3) |
| SC_2.5 | None | 0.6 | 2.5 | 2.5 | 3 | 3 | 970 | 85 | Not tested |
| SC_3 | None | 0.6 | 3 | 3 | 3 | 3 | 1482 | 89 | xy (1) z (1) |
| SC_3.5 | None | 0.6 | 3.5 | 3.5 | 3 | 3 | 2147 | 91 | Not tested |
| SC_4 | None | 0.6 | 4 | 4 | 2 | 2 | 941 | 91 | xy (3) z (3) |
| ST_1_0.6 | None | 0.6 | 1 | 0.6 | 6 | 8 | 1061 | 57 | xy (3) z (3) |
| ST_1_0.8 | None | 0.6 | 1 | 0.8 | 6 | 7 | 1082 | 61 | Not tested |
| ST_1.5_0.6 | None | 0.6 | 1.5 | 0.6 | 5 | 8 | 1257 | 66 | xy (1) z (1) |
| ST_1.5_0.8 | None | 0.6 | 1.5 | 0.8 | 5 | 7 | 1281 | 69 | Not tested |
| ST_1.5_1 | None | 0.6 | 1.5 | 1 | 5 | 6 | 1257 | 71 | Not tested |
| ST_2_0.6 | None | 0.6 | 2 | 0.6 | 4 | 8 | 1234 | 72 | xy (3) z (3) |
| ST_2_0.8 | None | 0.6 | 2 | 0.8 | 4 | 7 | 1258 | 74 | Not tested |
| ST_2_1 | None | 0.6 | 2 | 1 | 4 | 6 | 1234 | 76 | Not tested |
| ST_2_1.5 | None | 0.6 | 2 | 1.5 | 4 | 5 | 1343 | 80 | Not tested |
| ST_2.5_0.6 | None | 0.6 | 2.5 | 0.6 | 3 | 8 | 1000 | 75 | Not tested |
| ST_3_0.6 | None | 0.6 | 3 | 0.6 | 3 | 8 | 1326 | 78 | xy (1) z (1) |
| ST_3_0.8 | None | 0.6 | 3 | 0.8 | 3 | 7 | 1352 | 80 | Not tested |
| ST_3_1 | None | 0.6 | 3 | 1 | 3 | 6 | 1326 | 82 | Not tested |
| ST_3_1.5 | None | 0.6 | 3 | 1.5 | 3 | 5 | 1443 | 85 | Not tested |
| ST_3_2 | None | 0.6 | 3 | 2 | 3 | 4 | 1430 | 87 | Not tested |
| ST_3_2.5 | None | 0.6 | 3 | 2.5 | 3 | 3 | 1287 | 88 | Not tested |
| ST_3.5_0.6 | None | 0.6 | 3.5 | 0.6 | 3 | 8 | 1697 | 81 | Not tested |
| ST_4_0.6 | None | 0.6 | 4 | 0.6 | 2 | 8 | 980 | 81 | xy (3) z (3) |
| BCC_1.4 | Internal | 0.6 | 1.4 | 1.4 | 5 | 5 | 1191 | 54 | xy (1) z (1) |
| BCC_2 | Internal | 0.6 | 2 | 2 | 4 | 4 | 1331 | 68 | Not tested |
| BCC_2.4 | Internal | 0.6 | 2.4 | 2.4 | 3 | 3 | 885 | 74 | xy (1) z (1) |
| BCC_3 | Internal | 0.6 | 3 | 3 | 3 | 3 | 1482 | 81 | Not tested |
| BCC_3.4 | Internal | 0.6 | 3.4 | 3.4 | 3 | 3 | 2000 | 84 | xy (1) z (1) |
| FCC_1.4 | Wall | 0.6 | 1.4 | 1.4 | 5 | 5 | 1191 | 45 | xy (1) z (1) |
| FCC_2 | Wall | 0.6 | 2 | 2 | 4 | 4 | 1331 | 60 | Not tested |
| FCC_2.4 | Wall | 0.6 | 2.4 | 2.4 | 3 | 3 | 885 | 67 | xy (1) z (1) |
| FCC_3 | Wall | 0.6 | 3 | 3 | 3 | 3 | 1482 | 75 | Not tested |
| FCC_3.4 | Wall | 0.6 | 3.4 | 3.4 | 3 | 3 | 2000 | 79 | xy (1) z (1) |

$N$ and $M$ define the number of holes along xy and z, respectively. In addition, $V_{v,th}$ is the theoretical void volume fraction and column "Exp. test" reports the loading direction and the number of specimens tested

$$\rho_{rel} = \left(\frac{t}{w+t}\right)^2 \left(1 + \frac{2w}{h+t}\right) \quad \text{(Eq 6)}$$

In the specific case of SC cells, $h$ is equal to $w$, so that Eqs 5 and 6 become, respectively:

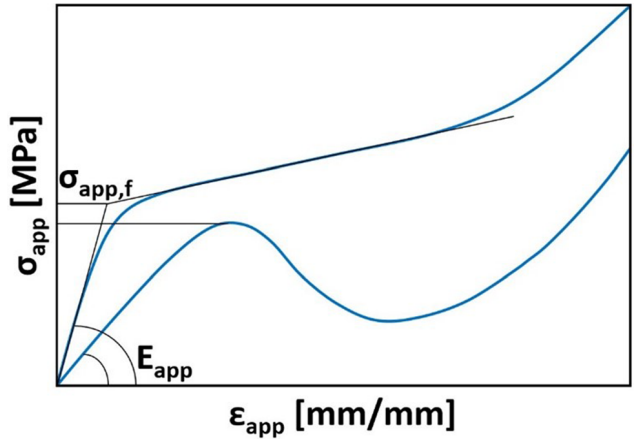

**Fig. 2** Schematic representation of two representative apparent stress, $\sigma_{app}$, vs. apparent strain, $\varepsilon_{app}$, curves, obtained from the compression tests. The computation of the compressive apparent modulus, $E_{app}$, and of the compressive apparent failure stress, $\sigma_{app,f}$, is shown for each curve (one in which $\sigma_{app,f}$ is denoted as "knee" of the curve, one in which $\sigma_{app,f}$ is denoted as a maximum point)

$$\frac{E_{ROM}}{E_c} = \frac{\sigma_{ROM,f}}{\sigma_{c,f}} = \frac{t^2}{(w+t)^2} \quad \text{along } z \text{ and } xy \tag{Eq 7}$$

$$\rho_{rel} = 3\left(\frac{t}{w+t}\right)^2 - 2\left(\frac{t}{w+t}\right)^3 \tag{Eq 8}$$

Finally, substitution of Eq. 7 into Eq. 8 results in the following analytical model for SC cells:

$$\rho_{rel} = 3\frac{E_{ROM}}{E_c} - 2\left(\frac{E_{ROM}}{E_c}\right)^{\frac{3}{2}} = 3\frac{\sigma_{ROM,f}}{\sigma_{c,f}} - 2\left(\frac{\sigma_{ROM,f}}{\sigma_{c,f}}\right)^{\frac{3}{2}} \tag{Eq 9}$$

In the specific case of ST cells having $h = 0.6$ mm $= t$, Eqs 5 and 6 become, respectively:

$$\frac{E_{ROM}}{E_c} = \frac{\sigma_{ROM,f}}{\sigma_{c,f}} = \begin{cases} \frac{t^2}{(w+t)^2} & \text{along } z \\ \frac{t}{2(w+t)} & \text{along } xy \end{cases} \tag{Eq 10}$$

$$\rho_{rel} = \frac{t}{w+t} \tag{Eq 11}$$

Finally, substitution of Eq. 11 into Eq. 10 results in the following analytical model for ST cells:

$$\frac{E_{ROM}}{E_c} = \frac{\sigma_{ROM,f}}{\sigma_{c,f}} = \begin{cases} \rho_{rel}^2 & \text{along } z \\ 0.5\rho_{rel} & \text{along } xy \end{cases} \tag{Eq 12}$$

**4.1.2 Analytical Equations for BCC and FCC Unit Cells.** Considering BCC and FCC unit cells, simple relative density expressions cannot be found because of the complexity of calculating the exact volume occupied by their struts. Therefore, such volume is extracted from CAD models using Solidworks, and $\rho_{rel}$ is separately calculated for each lattice structure.

Moreover, when applying Eqs 3 and 4 to calculate the mechanical properties of FCC and BCC lattices, the number of struts parallel to the loading direction is corrected in the attempt of considering also the contribution of diagonal struts. In particular, each strut parallel to the loading direction counts as 1, while diagonal struts are decomposed along the axes x, y, and z as if they were vectors (Fig. 3), so that the total value of $V_{//}$ becomes:

$$V_{//} = \begin{cases} \frac{\pi}{4}\left(\frac{t}{A}\right)^2 \left[(N+1)^2 + \frac{4}{\sqrt{3}}N^2\right] & \text{for BCC lattices} \\ \frac{\pi}{4}\left(\frac{t}{A}\right)^2 \left[(N+1)^2 + \frac{4}{\sqrt{2}}N(N+1)\right] & \text{for FCC lattices} \end{cases} \tag{Eq 13}$$

### 4.2 Semi-Empirical Model

In the second approach, the contribution of all struts to the stiffness and strength of the lattices is taken into account, in order to compensate for possible underestimations in the former approach, which considers only the contribution of struts parallel to the loading direction.

Considering the complexity of the geometry of the additional struts (being composed of multiple 3D-printed filaments and providing additional constraints in multiple directions), a semi-empirical model is developed.

The stiffness and the strength of the lattices are calculated as the sum of two contributions: (i) the contribution of struts parallel to the loading directions, which is again described according to the ROM (Eqs 3 and 4); (ii) the additional contribution ($\Delta E$ or $\Delta \sigma_f$) of other, non-parallel, struts, which is expressed as a function of their volume fraction ($V_{add}$). They result in the following equations:

$$E = E_{ROM} + \Delta E = E_c(V_{//} + K_0 + K_1 V_{add}) \tag{Eq 14}$$

$$\sigma_f = \sigma_{ROM,f} + \Delta \sigma = \sigma_{c,f}(V_{//} + K_0 + K_1 V_{add}) \tag{Eq 15}$$

with

$$\frac{\Delta E}{E_c} = \frac{\Delta \sigma_f}{\sigma_{c,f}} = K_0 + K_1 V_{add} \tag{Eq 16}$$

In Eqs 14-16, $V_{add} = 1 - V_{v,exp} - V_{//}$, while the values of constants $K_0$ and $K_1$ are derived as best fit parameters from experimental data on PLA lattices, more specifically from $\Delta E/E_c$ versus $V_{add}$ plots for each lattice family. Based on Eq. 14, $\Delta E$ is calculated by subtracting $E_{ROM}$ from the experimental value of the modulus $E$. Each $\Delta E/E_c$ versus $V_{add}$ plot is fitted with a linear equation (Eq. 16), in which the fitting constant values are obtained by linear regression; in particular, $K_0$ corresponds to the intercept of the line and the constant $K_1$ corresponds to its slope.

### 4.3 Artificial Neural Network

The third approach consists in the development of a single ANN model, as summarized in the following subsections. The modeling and implementation of the network are conducted using a code developed in the neural network environment of the software MATLAB (MathWorks®, Natick, MA, USA).

**4.3.1 Data Pre-Processing.** The neural network takes as inputs the parameters defining the geometry of the lattice structure and relates them to the output variables describing the mechanical properties of interest. In particular, the inputs are the hole width ($w$), the hole height ($h$), the number of holes in x (or y) direction ($N$), the number of holes in z direction ($M$), and the type of diagonal struts (none for SC and ST cells, internal

Journal of Materials Engineering and Performance

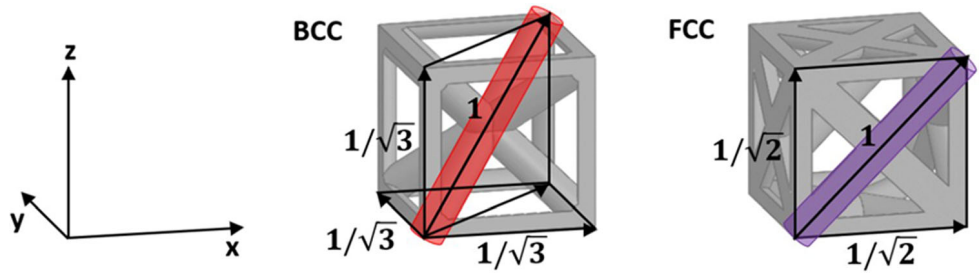

**Fig. 3** Decomposition of diagonal struts along the axes x, y, and z in BCC and FCC lattice structures

**Table 2** One-hot encoding adopted for representing the type of diagonal

| No diagonal | Internal diagonal | Wall diagonal |
|---|---|---|
| 1 | 0 | 0 |
| 0 | 1 | 0 |
| 0 | 0 | 1 |

**Table 3** Label encoding adopted for representing the type of diagonal

| Type of diagonal | Adopted label |
|---|---|
| No diagonal | 0 |
| Internal diagonal | 1 |
| Wall diagonal | 2 |

diagonal for BCC cells, and wall diagonal for FCC cells); the outputs are the apparent moduli ($E_{app,xy}$ and $E_{app,z}$) and the failure stresses ($\sigma_{app,f,xy}$ and $\sigma_{app,f,z}$).

Among the chosen inputs, a categorical variable is considered to describe the three types of diagonals, i.e., none ($D_1$), internal ($D_2$), and wall ($D_3$). Therefore, it is essential to convert this variable into a numerical format. Two commonly used techniques for encoding categorical variables are one-hot and label encoding (Ref 61). In most scenarios, one-hot encoding is the preferred way to convert a categorical variable into a numeric variable because label encoding makes it seem that there is a ranking between values. However, one drawback of one-hot encoding is that it requires as many new variables as there are unique values in the original categorical variable. Both techniques are considered and compared here by using the representation reported in Tables 2 and 3. However, in the present case, no evident differences in terms of results' accuracy are noted between the two techniques and, from here on, the one-hot encoding is maintained.

Finally, the input and output variables have a wide span and should be normalized and scaled appropriately to improve the efficiency of the learning process and the prediction capability of the proposed network (Ref 62). A normalization between 0 and 1 is found to be effective in this work and is thus implemented exploiting the Matlab built-in function 'normalize'.

**4.3.2 Network Setup and Hyperparameter Analysis.** The overall architecture of the proposed network is shown in Fig. 4.

The configuration and accuracy of the network is optimized in terms of the following hyperparameters:

1. *Number of hidden layers.* This item, together with the second one, determines the complexity of the neural network. Theoretically, increasing the number of hidden layers improves the ability of the neural network to tackle nonlinear problems, but the learning process becomes more difficult and time-consuming. Thus, with this increase, the prediction accuracy generally increases first and then decreases, resulting in overfitting issues. The use of a network with one hidden layer (shallow network) generally gives better performance than multi-hidden-layer network (deep network) when dealing with small datasets (Ref 63), such as the case being considered here. A single hidden layer of neurons is thus used between the inputs and the outputs.

2. *Number of nodes in the hidden layers.* Following the discussion provided for the previous item, two nodes have been adopted for the hidden layer.

3. *Activation function in the hidden nodes to process the inputs.* The activation function ensures that the neural network is qualified for modeling the nonlinear phenomena, and the learning process can be carried out efficiently. Selecting the activation function should be achieved according to the specific research object. In this work, for predicting the mechanical properties whose values are always positive, the 'logsig' function available in the MATLAB environment is adopted.

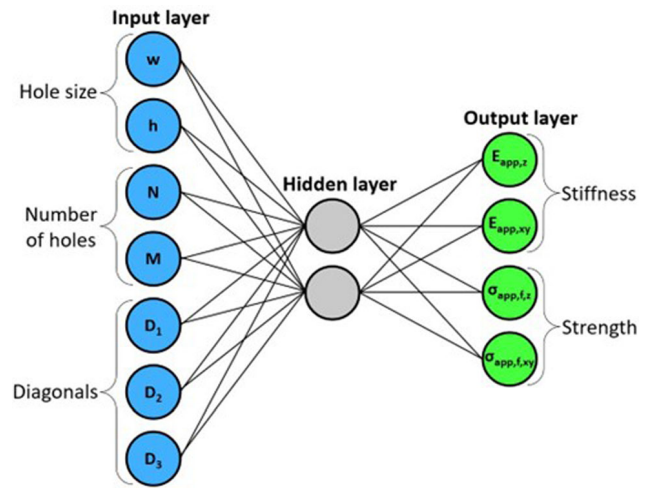

**Fig. 4** Overall architecture of the network

4. *Back-propagation learning algorithm.* It is used to optimally compute the internal parameters of the network (i.e., weights and biases). Basically, the weights and biases are initially set randomly and then updated iteratively by calculating the error on the training outputs and distributing it back to the layers (Ref 64). Here, Bayesian regularization backpropagation ('trainbr' in the MATLAB environment) is adopted; this learning algorithm updates the values of weights and biases according to Levenberg-Marquardt optimization and minimizes a linear combination of squared errors and weights, so as to produce a network with good generalization qualities (Ref 65, 66).

### 4.3.3 Learning Process.
The data are divided into three groups: (i) training data, (ii) validation data, and (iii) test data.

Data are split into 70/20/10 proportions. Other proportions (70/10/20, 70/15/15) are used and compared.

70% of the experimental results are chosen for training since reducing the training data is likely to give a worse predictive network, as the network is not enough comprehensively trained. The training set is used for updating the network weights and biases through the backpropagation learning algorithm.

20% of the results are used for validation. This phase is important, as it explores how good the solution is at predicting the results on data that have not been used for training. If the validation error is poor, the parameters used for training the network are adjusted to improve the results for the validation data set. In fact, if these parameters have not been set correctly, it is possible that the training data have been 'over-fitted', resulting in poor predictions on unseen data. The validation error normally decreases during the initial phase of training, as does the training set error. However, when the network begins to overfit the data, the error on the validation set typically begins to rise. The network weights and biases are saved at the minimum of the validation set error. The reader is referred to (Ref 67-69) for a detailed discussion on possible strategies to overcome overfitting in the case of small datasets.

Test data (10% of the experimental data) are kept until the overall network has been optimized and are used to assess how good the network is at predicting the results for entirely unseen data. It is important that the network is not changed further as a result of examining performance on the test data, as this may bias the network for future unseen data. If the error on the test set reaches a minimum at a significantly different iteration number than the validation set error, this might indicate a poor division of the data set.

The function 'divideint', provided in the MATLAB environment and based on using interleaved indices, is employed for dividing the data into training, validation, and test sets.

The performance of the network can be assessed by means of commonly used loss functions in MATLAB (e.g., mean relative error, mean absolute error, mean square error). This work adopts the mean square error "mse", which measures the network's performance according to the mean of squared errors between the network outputs and the experimental data.

### 4.4 Finite Element Modeling

The fourth and last approach consists in simulations of compression tests by means of FEA using the commercial finite element software Abaqus 2023 (Simulia, Providence, RI, USA).

Lattice structures labeled as SC and ST (Fig. 1) are meshed using eight-node brick elements (C3D8), while structures labeled as BCC and FCC (Fig. 1) are meshed using four-node tetrahedral elements (C3D4), both available in the software library. A mesh convergence study is performed and an example of adopted meshes is reported in Fig. 5.

Since the FEA approach is here investigated for elastic moduli predictions, an isotropic linear elastic constitutive model is adopted. Model parameters are a Young's modulus equal to 2000 MPa, identified by numerical simulation on the basis of apparent stress-strain curves obtained for the 3D-printed PLA structure SC_3, and a Poisson's ratio equal to 0.35, which is a reasonable value for thermoplastic polymers and is consistent with literature data on PLA filaments (Ref 70).

To simulate the mechanical tests on lattice structures under uniaxial compression, a quasi-static analysis is performed. A displacement, whose value is equal to the experimental one (taken before the "knee" or the peak in the stress-strain curve), is applied perpendicularly to a face of the structure, along the $x$ or $z$ direction (y is equivalent to x), while the opposite face is constrained along the three axes.

Apparent stresses and strains are computed from the displacement vs reaction curve, following the definitions provided in Sect. 3.3.

### 4.5 Comparison Between the Approaches

To quantitatively compare the accuracy of the four approaches, it is here proposed to use the coefficient of determination, $R^2$, corresponding to the linear association between experimental and computed properties. $R^2$ is defined as in standard statistical textbooks, as follows:

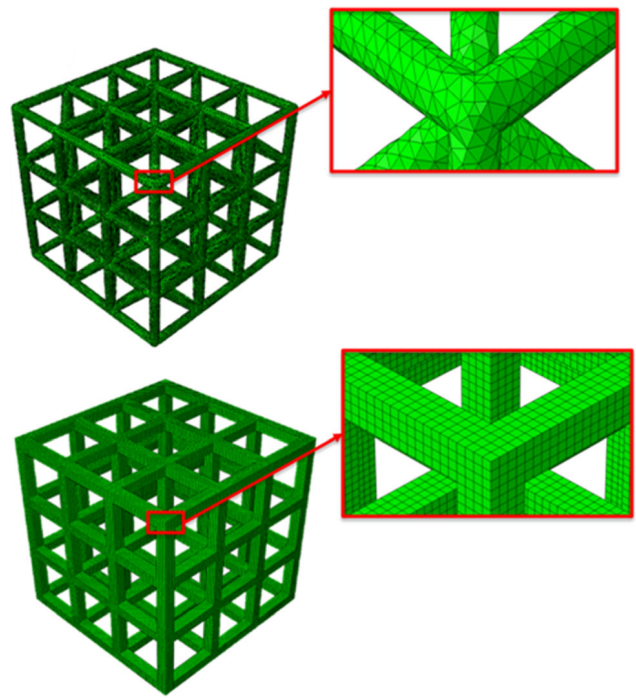

**Fig. 5** Examples of meshes adopted for (a) SC and ST lattices, and for (b) BCC and FCC lattices

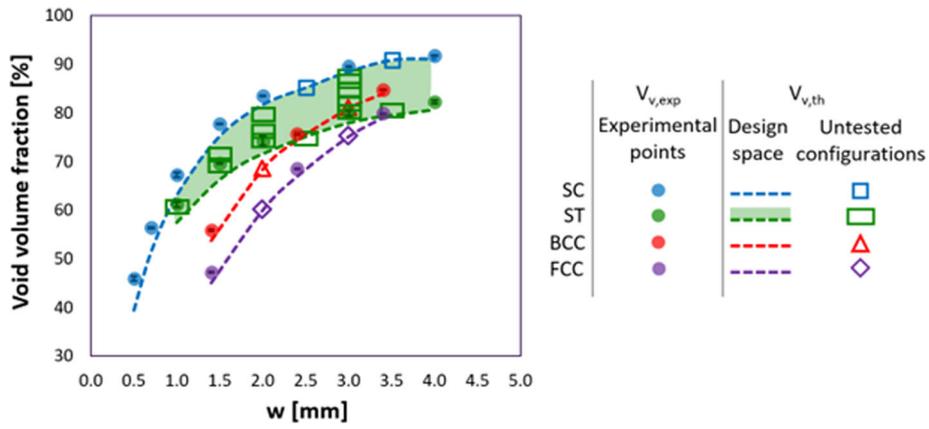

**Fig. 6** Experimental and theoretical values of void volume fraction ($V_{v,\exp}$ and $V_{v,\text{th}}$, respectively) plotted against hole width ($w$) for PLA lattices with SC, ST, BCC, and FCC unit cells

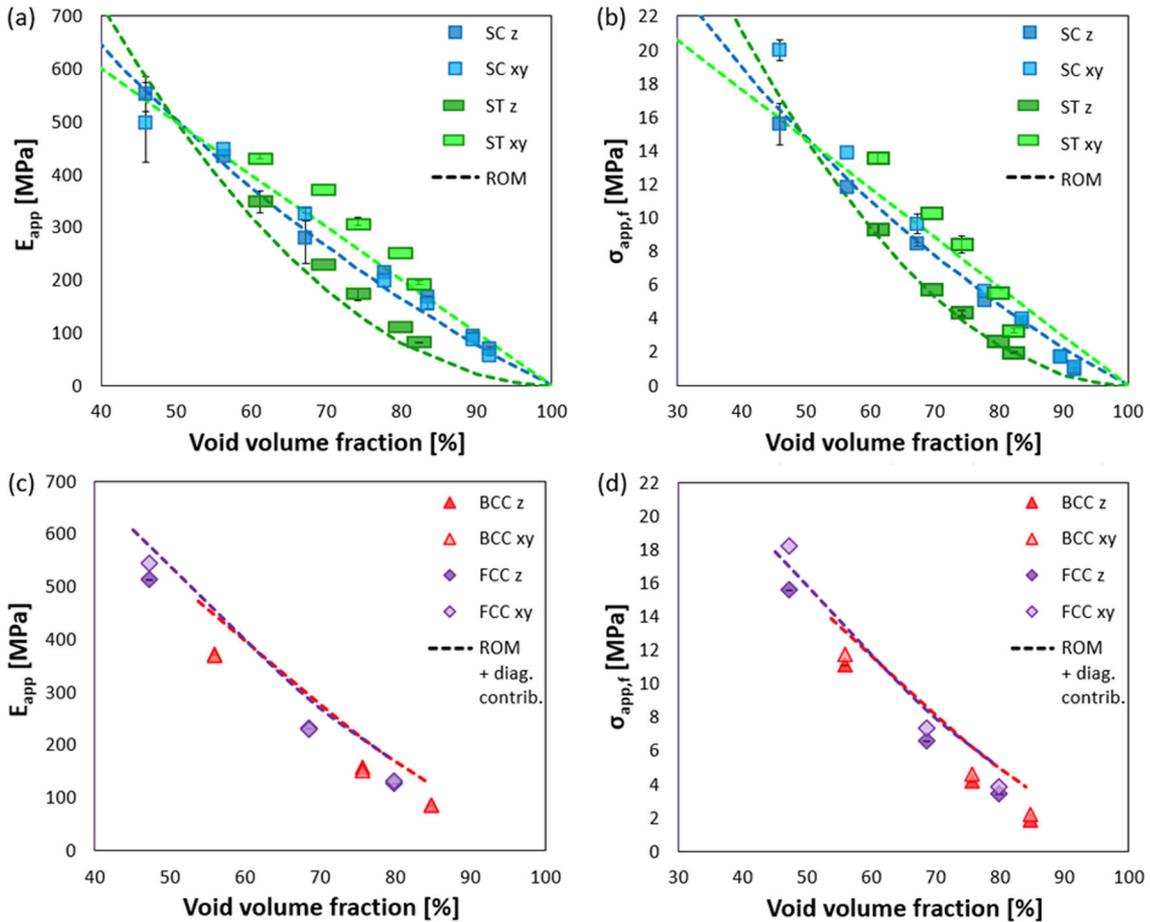

**Fig. 7** Experimental values and predictions of the analytical model based on the rule of mixtures (ROM) for apparent modulus (a, c) and failure stress (b, d) plotted against void volume fraction, for PLA lattices with SC and ST unit cells (a, b) and with BCC and FCC unit cells (c, d)

$$R^2 = 1 - \frac{\sum_j \left(v_j^{\exp} - v_j^{\text{mod}}\right)^2}{\sum_j \left(v_j^{\exp} - \bar{v}\right)^2} \quad \text{(Eq 17)}$$

where $v_j$ is the mechanical property related to experiment $j$ (superscripts exp and mod stand for experimental and computed values, respectively), and $\bar{v}$ is the mean of the experimental data. Accordingly, for each approach four $R^2$ have been computed, corresponding to the four mechanical properties measured (i.e., the apparent moduli and the failure stresses in xy and z directions). For the approach based on FEA, only two $R^2$ have been computed for the apparent moduli. It is also noted that in the case of BCC and FCC structures, just one

FEA was conducted and thus $j = 1$ in Eq. 17. In such case the coefficient is simply computed using the relative error, as $1 - |v^{\text{exp}} - v^{\text{mod}}|/v^{\text{exp}}$.

## 5. Results and Discussion

This section presents the results of the adopted modeling approaches, discussing their ability to match the experimental campaign results and to predict the mechanical properties of new untested configurations.

### 5.1 Experimental Results

The details of the geometry of SC, ST, BCC, and FCC lattice structures under investigation are provided in Sect. 2 (see Fig. 1 and Table 1). Cubic specimens were printed by FDM, obtaining structures with a wide range of void volume fraction values, depending both on the unit cell dimensions and on the strut arrangement. As the hole width ($w$) increases (between 0.5 mm and 4 mm), the void volume fraction ranges from about 45% to more than 90%, as shown in Fig. 6. The graph also shows that experimental void volume fractions ($V_{v,\text{exp}}$) are consistent with the corresponding theoretical values ($V_{v,\text{th}}$), with small deviations of about 1-5% that tend to disappear as the cell size grows. In addition, $V_{v,\text{th}}$ values for untested lattice configurations are displayed, covering unoccupied regions in the design space, which is delimited by the lattices with the smallest and biggest holes in each series; note that the design space for ST lattices extends until reaching that of SC lattices when the hole height and width are equal ($h = w$).

The mechanical properties of PLA lattices were evaluated by compression tests in terms of stiffness (apparent modulus) and strength (apparent failure stress), as described in Sect. 3. It must be taken into account that all the lattices have an anisotropic structure due to their layer-by-layer fabrication, resulting in a necessary distinction between their build direction ($z$) and the other two directions ($x$ and $y$) forming the cartesian coordinate system in Fig. 1; therefore, tests were performed along both z and $xy$ directions. The resulting stiffness and strength values are plotted against the void volume fraction and compared with those of the different modeling approaches in Fig. 7 (analytical model), Fig. 9 (semi-empirical model), and Figs. 12 and 13 (ANN and FEA). Both stiffness and strength are observed to decrease as the void volume fraction increases, varying over about one order of magnitude (apparent modulus between about 50 and 550 MPa and apparent failure stress between about 1 and 20 MPa). Lattices with cubic cells (SC, BCC, and FCC) have similar stiffness along all directions, while their compressive strength tends to be slightly higher along the $x$ (or $y$) direction; conversely, lattices with tetragonal cells (ST) are highly anisotropic because their hole height ($h$) is lower than their hole width ($w$), resulting in a higher number of struts along x (or y) and, as a consequence, significantly higher $E_{app,xy}$ and $\sigma_{app,f,xy}$. For the same value of void volume fraction, different values of stiffness and strength are found depending on the lattice unit cell and the loading direction, highlighting the need for different equations to describe the mechanical properties corresponding to different lattice types and directions. However, for SC, BCC, and FCC structures, given the modest influence of the loading direction on their properties, the same analytical and semi-empirical equations were employed for their values of $E_{app,z}$ (or $\sigma_{app,f,z}$) and $E_{app,xy}$ (or $\sigma_{app,f,xy}$); instead, ANN and FEA strategies considered the different directions separately.

### 5.2 Analytical and Semi-Empirical Model Results

The first two approaches here presented are developed for strut-based lattice types and they are based on the rule of mixtures (ROM) typically employed in composite materials, considering lattice struts parallel to the loading direction as fibers and the voids as matrix. In the beginning, a "primitive" purely analytical version is described, and it then sets the bases for deriving a more accurate semi-empirical model.

As a first approach toward modeling of PLA lattice structure-property correlations, simpler analytical equations are investigated. As described in Sect. 4.1, these equations are based on the ROM, considering only lattice struts parallel to the loading direction as if they were the composite fibers and neglecting perpendicular struts; as for diagonal struts, they contribute only with their parallel component. Interestingly, this modeling approach leads to equations (see Eqs 9 and 12) that resemble Gibson-Ashby power law relationship between relative stiffness/strength and relative density of cellular structures (Ref 1, 5, 19, 32-43, 58), with different values of the exponent (between 1 and 2) depending on the strut arrangement. In its simplicity, this analytical model is able to capture the qualitative trend of apparent modulus and failure stress plotted against the lattice void volume fraction (Fig. 7), but it fails in providing good quantitative predictions of the mechanical properties. In

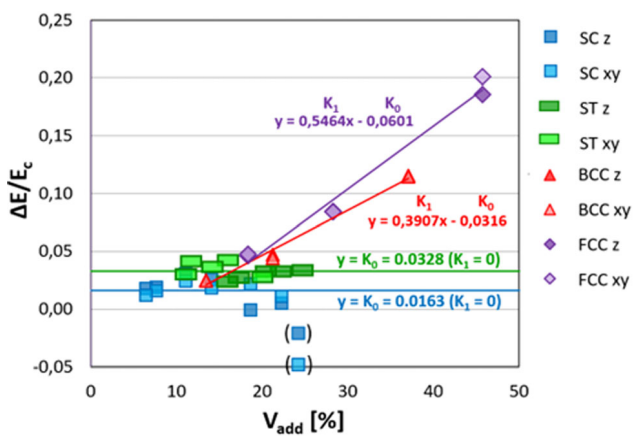

**Fig. 8** Graph used for the determination of the fitting constants $K_0$ (intercept) and $K_1$ (slope) for the semi-empirical model of PLA lattice structure-property correlations: the contribution of additional struts to the modulus ($\Delta E$) is normalized with respect to PLA modulus ($E_c$) and plotted against the volume fraction of additional struts ($V_{add}$); data in brackets are excluded from the fitting

**Table 4** Fitting constants $K_0$ and $K_1$ derived from $\Delta E/E_c$ versus $V_{add}$ plots for PLA lattice structures (Fig. 7).

| Lattice unit cell | $K_0$, - | $K_1$, - |
|---|---|---|
| SC | 0.0163 | 0 |
| ST | 0.0328 | 0 |
| BCC | − 0.0316 | 0.3907 |
| FCC | − 0.0601 | 0.5464 |

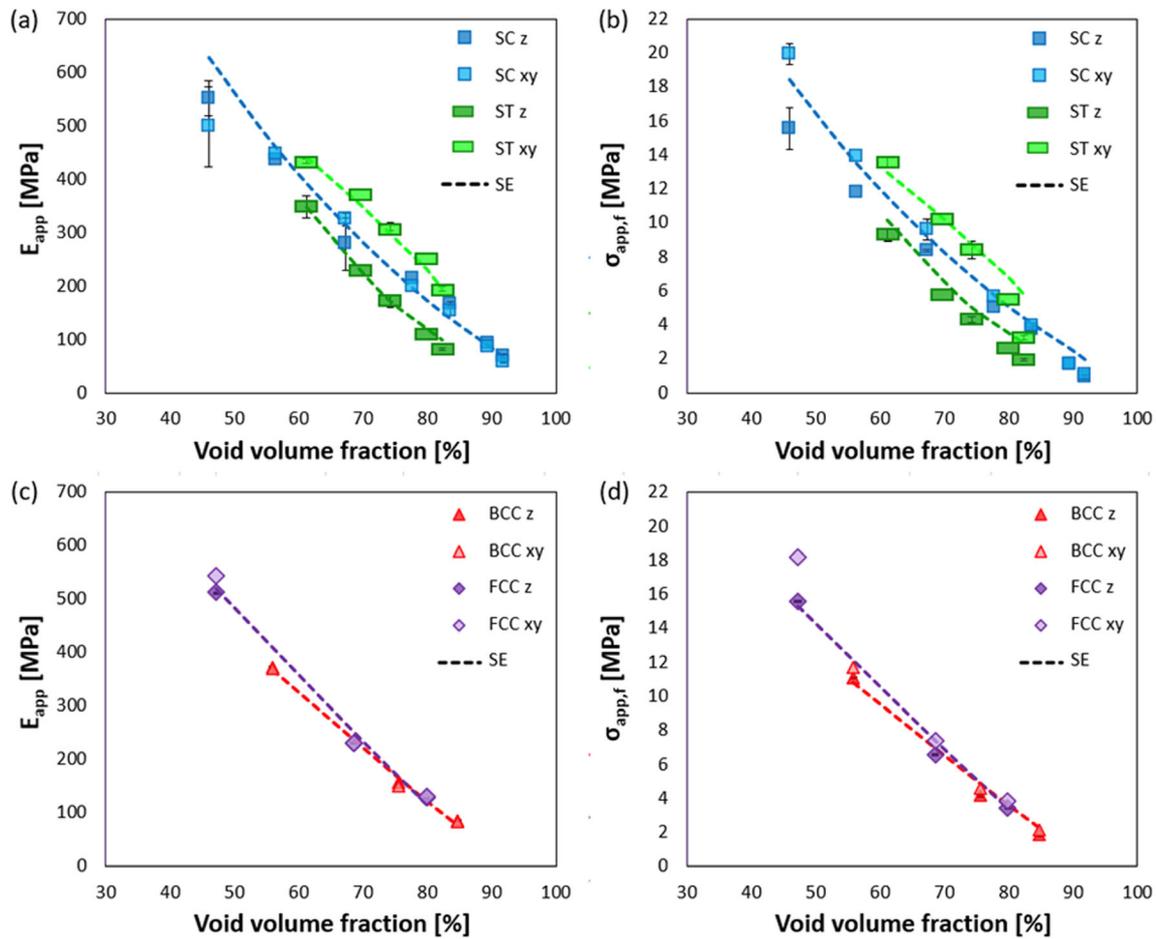

**Fig. 9** Experimental values and semi-empirical model predictions (SE) of apparent modulus (a, c) and failure stress (b, d) plotted against void volume fraction, for PLA lattices with SC and ST unit cells (a, b) and with BCC and FCC unit cells (c, d)

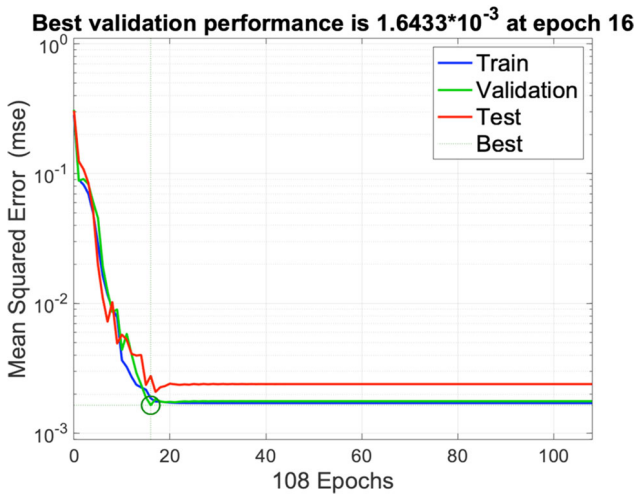

**Fig. 10** Mean squared error ("mse") values obtained along the epochs for training, validation, and test phases. They are calculated for both stiffness and strength ANN outputs with respect to the corresponding experimental targets

fact, stiffness (Fig. 7a) and strength (Fig. 7b) of SC and ST lattices tend to be underestimated with respect to their experimental values, suggesting that the constraint provided by struts perpendicular to the loading direction is not negligible and improves the lattice performance to some extent. Conversely, stiffness (Fig. 7c) and strength (Fig. 7d) of BCC and FCC lattices are overestimated by the analytical model, which may be due to the simplicity of the assumption that the contribution of diagonal struts can be decomposed into parallel (contributing) and perpendicular (not contributing) components, but could also be ascribed to the anisotropic nature of FDM 3D-printed materials; in fact, the interface between layers is a potential source of weakness, that appeared mostly negligible under compression, but may reduce the performance of diagonal struts, in which the layers are not orthogonal with respect to the strut axis. More in general, the difficulty of developing equations that can accurately take into account all geometrical details of the lattice structures is a major issue of purely analytical approaches. In fact, it is not trivial to take into account the constraint played by struts perpendicular/inclined with respect to the load, the geometry of the volumes defined by the intersection of multiple cylindrical struts, as well as the nonhomogeneous and anisotropic microstructure obtained by FDM. In light of such limitations of the analytical model, a semi-empirical model is developed as described in Sect 4.2. Like the Gibson-Ashby approach, this model requires fitting with experimental data to define the values of fitting parameters but maintains the contributions of struts with different orientations separate. As said, the mechanical properties of PLA

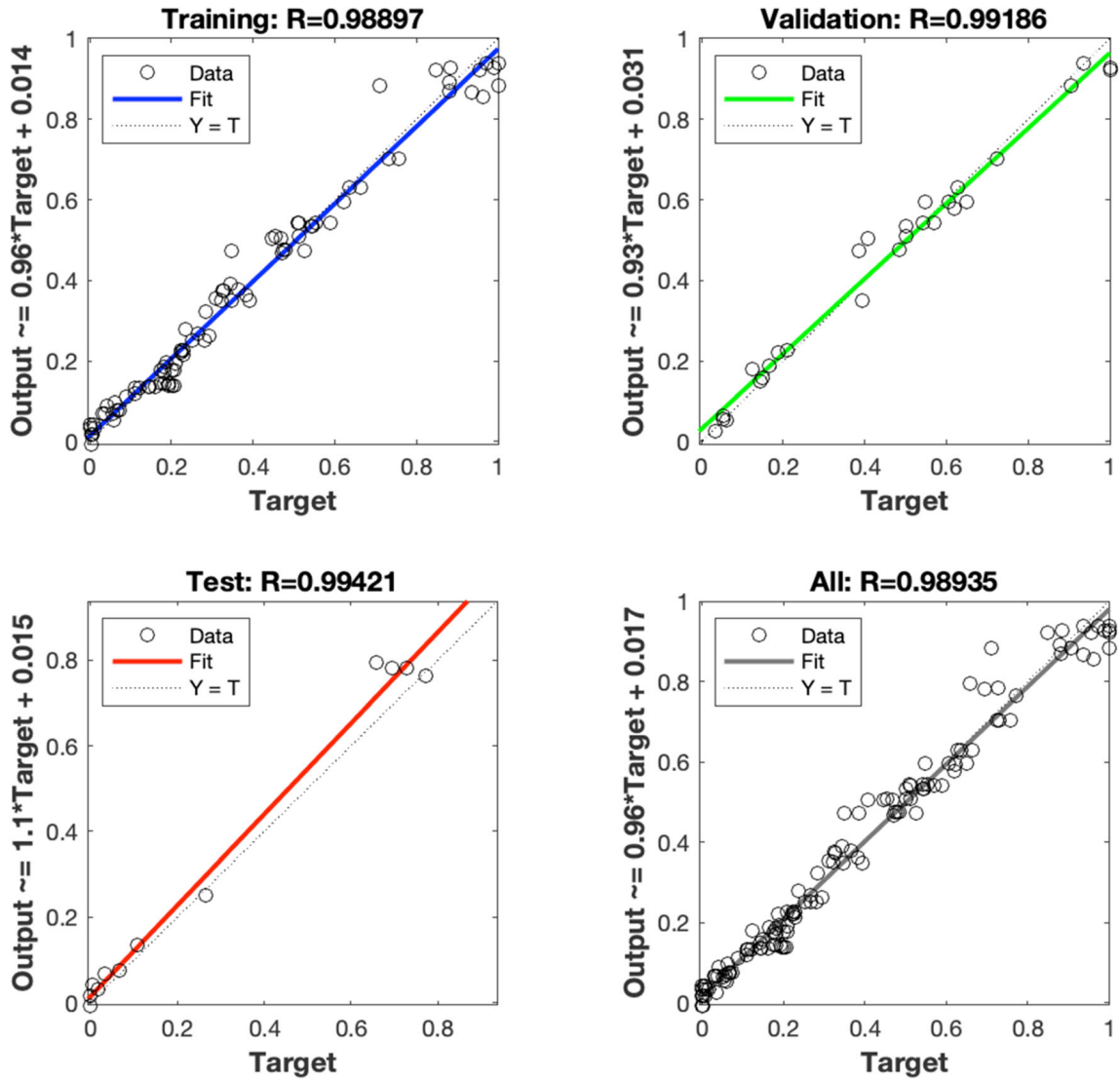

**Fig. 11** Correlation of ANN outputs with experimental target values for training, validation and test datasets, and for all datapoints considered together; normalized values of both apparent stiffness and apparent strength of PLA lattices are considered as output/target

lattices are expressed as the sum of two contributions: (i) $E_{ROM}$ or $\sigma_{ROM,f}$, which is a purely analytic component, associated to struts parallel to the loading direction and derived according to the ROM; (ii) $\Delta E$ or $\Delta \sigma_f$, which is experimentally measured and associated to the additional struts. $\Delta E$ can be normalized with respect to PLA stiffness ($E_c$) and plotted as a function of the volume fraction of additional struts ($V_{add}$), as shown in Fig. 8. This plot clearly highlights two possible responses from the lattice structures: (i) in the case of SC and ST unit cells, additional struts provide a moderate contribution to the lattice stiffness, which is apparently not influenced by the volume occupied, at least in the range of $V_{add}$ values here covered; (ii) in the case of BCC and FCC unit cells, thanks to diagonal struts, the lattice stiffness receives a higher additional contribution, which seems to increase linearly with $V_{add}$. The plot can be used to determine the values of the constants $K_0$ and $K_1$ in the semi-empirical model, according to Eq. 16: $K_0$ corresponds to the intercept, while $K_1$ corresponds to the slope of the experimental trends; data points between brackets correspond to exceptionally dense SC lattice structures ($w = 0.5$ mm $< t$),  which fall out of the trend and can not be included in the analysis. The two fitting constants assume specific values for the various lattice families, as reported in Table 4.

In SC and ST lattices, the addition of struts perpendicular to the loading direction is found to give a small positive contribution ($K_0 > 0$, corresponding to about 2-3% of 3D-printed PLA Young's modulus), independently of the volume occupied (i.e., $K_1 = 0$). Conversely, in BCC and FCC lattices, $\Delta E$ is proportional to the volume fraction of additional struts ($K_1 > 0$), and a negative intercept results from the fitting ($K_0 < 0$). In fact, mechanical properties increase significantly with the number of diagonal struts having a component parallel to the loading direction, but additional struts in BCC and FCC lattices also include several struts that are perpendicular to the load and contribute less; therefore, the negative value of $K_0$ compensates for these struts. Moreover, as expected, the constant $K_1$ is higher for FCC lattices than for BCC lattices, because wall diagonals are less inclined than internal diagonals with respect to the loading direction. From all these considerations, it is clear that the introduced constants are not simply

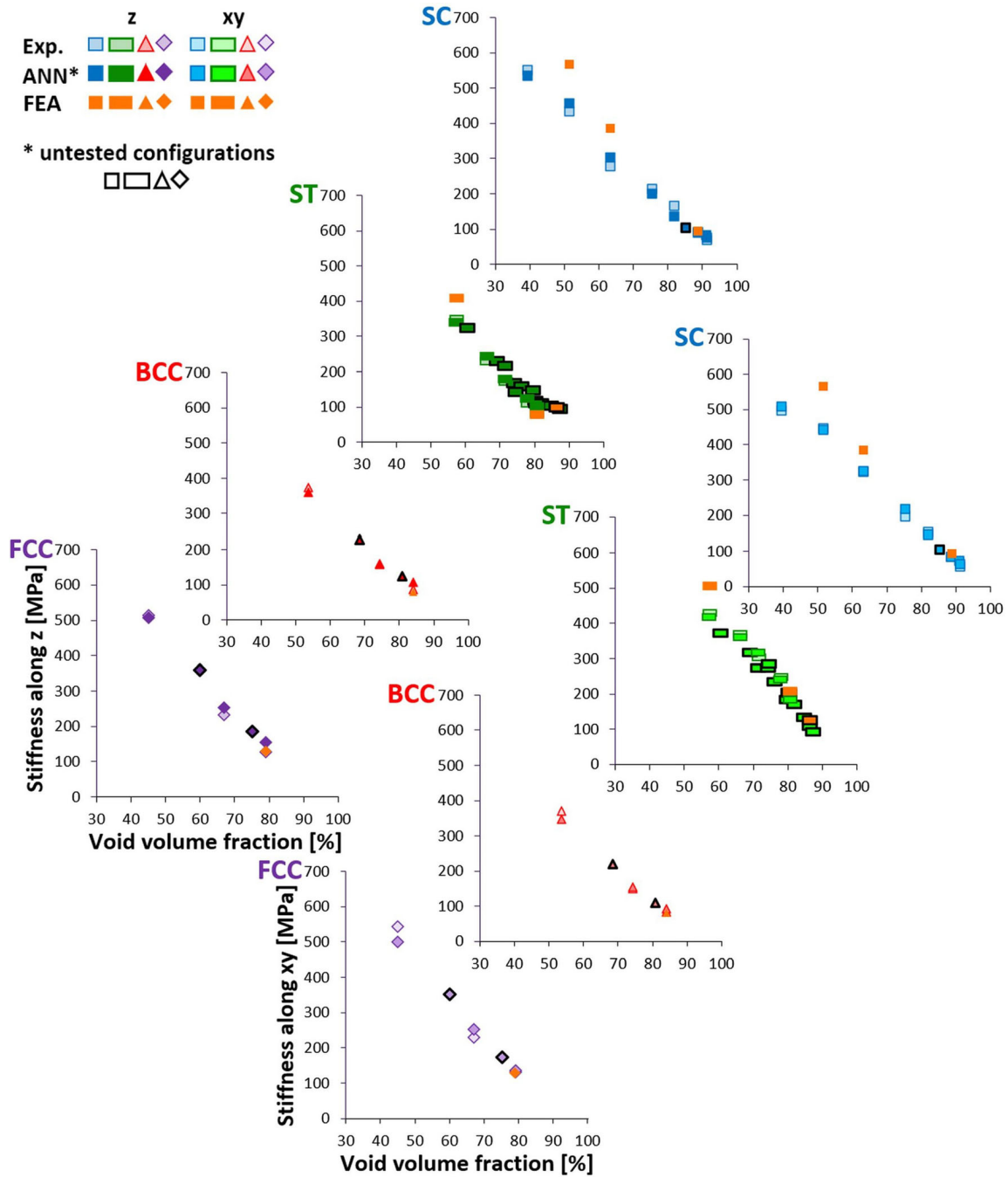

**Fig. 12** PLA lattice apparent modulus along z and xy directions, according to experiments, artificial neural network (ANN), and finite element analyses (FEA)

"fudge" factors to produce better fits, but meaningful parameters to properly account for additional struts in the model. In particular, $K_0$ is useful for including the effect of struts perpendicular to the loading direction, which was neglected in the previous analytical model, while the experimental evaluation of $K_1$ allows for a better estimation of the contribution of diagonal struts.

As a result, semi-empirical model predictions are much closer to the experimental data on PLA lattice structures, as reported in Fig. 9. Noteworthily, even if $K_0$ and $K_1$ are here derived from stiffness values only, the model can describe not only apparent modulus trends (Fig. 9a, c), but also apparent failure stress trends (Fig. 9b, d).

### 5.3 ANN and FEA Results

The optimum configuration of the neural network is determined based on the mean squared error evaluated for stiffness and strength outputs with respect to the measured experimental values. The mean squared error ("mse") is reported in Fig. 10 along the epochs (one epoch corresponds to one cycle through the full dataset), for each phase of the learning process (training, validation, test). In particular, the parameters of the network are saved at the epoch corresponding to the best validation performance (mean squared error of $1.6433 \times 10^{-3}$, at epoch 16).

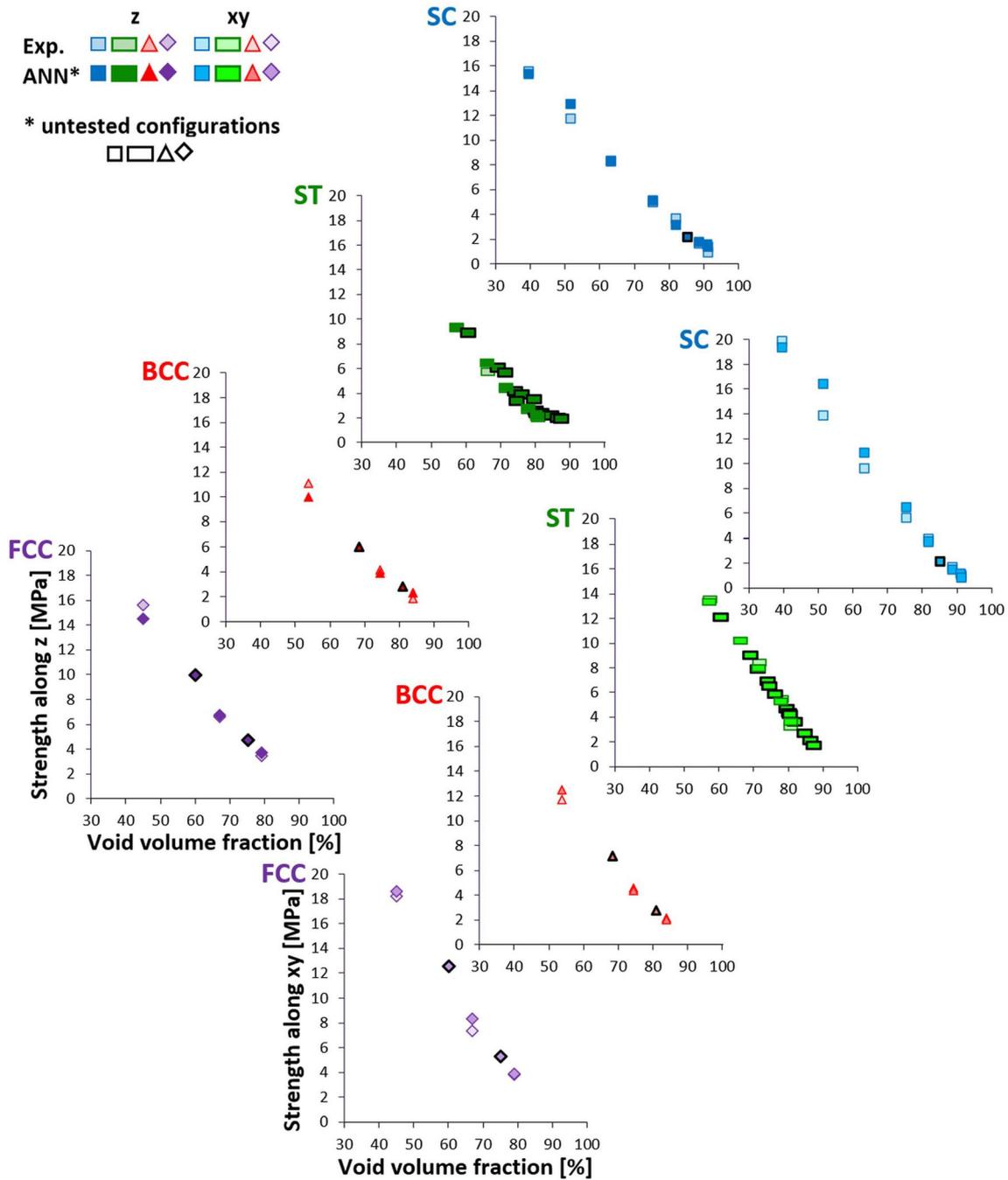

**Fig. 13** PLA lattice apparent failure stress along *z* and *xy* directions, according to experiments and artificial neural network (ANN)

In order to better visualize the overall accuracy of the predictions, the correlation between experimental and ANN results is shown in Fig. 11, together with the corresponding coefficients of correlation (*R*) computed directly in the neural network environment of the software MATLAB (through the built-in function 'corrcoef'). Normalized values of both apparent stiffness and apparent strength of PLA lattices are reported. The first three graphs refer to the training, validation, and test datasets, while the fourth one shows the results considering all 64 data points. The 45° line corresponds to perfect matching between predictions (output) and experimental values (target). The predictions are very good in terms of accuracy. In addition, since the points are distributed uniformly above and below the 45° line, it can be concluded that there is no over- or under-prediction tendency, and the prediction errors are fairly random (no systematic error is observed). It should be noticed here that the *R* coefficients for the validation and test datasets are higher than those for the training dataset.

The outputs of the ANN are reported in Fig. 12 (stiffness) and Fig. 13 (strength) for both tested and untested lattice configurations of Table 1, together with experimental data points and, for stiffness only, with the results of finite element analyses (FEA, highlighted with orange color). In particular, untested configurations correspond to data points surrounded by a black border. For each PLA lattice type (SC; ST; BCC; FCC), the four outputs ($E_{app,xy}$; $E_{app,z}$; $\sigma_{app,f,xy}$; $\sigma_{app,f,z}$) are plotted against the void volume fraction.

**Table 5**  Obtained coefficients of determination, $R^2$, for each approach and for each mechanical property

| Lattice unit cell | Load direction | $R^2_E$ Analytical model | $R^2_\sigma$ Analytical model | $R^2_E$ Semi-emp. model | $R^2_\sigma$ Semi-emp. model | $R^2_E$ ANN | $R^2_\sigma$ ANN | $R^2_E$ FEA | $R^2_\sigma$ FEA | Colors legend |
|---|---|---|---|---|---|---|---|---|---|---|
| SC | z | 0.9861 | 0.9873 | 0.9553 | 0.9199 | 0.9868 | 0.9892 | 0.5012 | n.a. | x>0.95 |
|  | xy | 0.9680 | 0.9382 | 0.9048 | 0.9852 | 0.9962 | 0.9682 | 0.7215 | n.a. | 0.9<x<0.95 |
| ST | z | 0.8466 | 0.9837 | 0.9899 | 0.8793 | 0.9839 | 0.9855 | 0.9030 | n.a. | 0.85<x<0.9 |
|  | xy | 0.6971 | 0.8290 | 0.9759 | 0.8676 | 0.9836 | 0.9926 | 0.7915 | n.a. | 0.8<x<0.85 |
| BCC | z | 0.6292 | 0.6175 | 0.9964 | 0.9835 | 0.9847 | 0.9671 | 0.9382 | n.a. | 0.75<x<0.8 |
|  | xy | 0.5980 | 0.7663 | 0.9940 | 0.9808 | 0.9870 | 0.9877 | 0.9536 | n.a. | 0.7<x<0.75 |
| FCC | z | 0.7845 | 0.8238 | 0.9923 | 0.9910 | 0.9852 | 0.9835 | 0.9872 | n.a. | 0.65<x<0.7 |
|  | xy | 0.8694 | 0.9575 | 0.9870 | 0.9236 | 0.9728 | 0.9903 | 0.9912 | n.a. | x<0.65 |

**Table 6**  Summary of the proposed and analyzed modeling approaches, together with their advantages and disadvantages

| Investigated model | Benefits | Drawbacks |
|---|---|---|
| Analytical model | Very simple; does not require experimental dataset | Neglects struts perpendicular to the load; poorly accounts for diagonal struts |
| Semi-empirical model | Good accuracy (based on real data); considers all struts contributions | Fails for denser structures; requires new experimental dataset if changing material/unit cell/strut thickness |
| ANN | Very good accuracy (based on real data); same approach for any cell type; reduced analysis time | Requires new experimental dataset if changing material/unit cell/strut thickness/build direction |
| FEA | Good accuracy; additional information on stress/strain field distribution; does not require experimental dataset | Poor predictions for denser structures; longer analysis and model preparation/meshing time; limited to the linear elastic field (may be extended at the cost of higher computational effort and modeling complexity) |

As highlighted above, the ANN is able to accurately estimate the mechanical properties of the lattices based on the inputs describing their structure. FEA results are also consistent with the experimental data, but larger deviations are observed for denser lattices (i.e., low void volume fractions). This may be due to the adopted constitutive model that may be too simple for the description of FDM 3D printed structures. More complex constitutive models as (Ref 71, 72) could be introduced to accurately model their anisotropic behavior and/or to predict the failure stress, but this at the cost of increased computational time and modeling complexity. As an alternative, data-driven constitutive models based on ANN could be employed, as proposed in (Ref 57).

Therefore, the data-driven approach is once again an effective approach. In particular, the ANN model is the most accurate among the proposed prediction strategies, being able to capture very well even the behavior of denser lattices, thanks to the training on real data. Finally, the network can also be used to evaluate the properties of untested PLA core structures (data points with black contour), which are found to be consistent with the overall trend of lattices having the same unit cell type. However, it is necessary to take into account that the proposed network is valid for structures with the same material, unit cell type, strut thickness, and build direction, as those investigated. Variations in these variables require new experimental data for the training. Further, it is worth noting that finite element simulations have been run on an AMD Ryzen 5 3500U with Radeon Vega Mobile Gfx 2.10 GHz processor, 800 GB 2400 MHz DDR4 RAM, 64-bit operating system, personal computer. Computational times vary between 45 s for the simulation of each ST structure and 532 s for each FCC structure. Thus, the conducted finite element analyses, although not heavy since dealing with linear elastic models, have computational times higher than that of the ANN approach, which requires few seconds for all the training/validation/test steps. Moreover, these times may increase if more complex models (discussed above) are adopted. Further, the construction of finite element geometry (i.e., the mesh) is costly, time-consuming and can lead to accuracy problems (Ref 73).

Table 5 accounts for the coefficients of determination $R^2$ computed for each approach. Results confirm the discussion provided above for each approach. In fact, the obtained values stretch in a reasonable range of reliability for almost all approaches. Highest values of $R^2$ for all cases are provided by the ANN, while lowest values of $R^2$ are attributed to the analytical model. The FEA model provides low values for the SC-based structures due to the above-mentioned limitations in simulating denser structures.

To conclude the discussion, Table 6 summarizes the advantages and disadvantages of the approaches analyzed and applied to the present strut-based lattice structures.

# 6. Conclusions

Several modeling approaches can be adopted for the prediction of the mechanical properties of lattice structures that could be employed in several application fields from medical and tissue engineering to automotive, mechanical, aeronautical, and seismic ones (Ref 2-4). This work has explored a selection of models belonging to different categories and applied them to the case of PLA lattice structures fabricated via FDM 3D printing, aiming to develop and compare helpful tools for their design.

The first approach presented, consisting of analytical equations based on the rule of mixtures, is very simple, and has the advantage of being purely theoretical. Therefore, it can be easily applied during the very first design stages to roughly estimate how the lattice properties vary with their architecture. However, it introduces significant approximations, especially neglecting the contribution of struts perpendicular to the loading direction, and cannot be considered quantitatively accurate.

Another strategy employing theoretical calculations relies on FEAs. These require higher computational effort to evaluate large numbers of geometrical configurations, but they are able to correctly consider the effect of all lattice struts and to provide additional information on stress and strain distributions. They may help optimizing lattices at later design stages, though the specific setting here investigated shows limitations in the analysis of denser lattices. It should be noted that this model does not consider the layered nature of 3D-printed specimens, which leaves space for improvement.

Due to the complex microstructure of PLA lattice structures realized by AM, building a reliable theoretical model can be particularly cumbersome, but it is possible to exploit a set of experimental results to create a predictive tool for new untested configurations. Indeed, the data-driven models here proposed performed very well, both in the case of semi-empirical mathematical equations and in the case of machine learning; the Gibson-Ashby model (Ref 1, 5, 19, 32-43, 58) also falls in the category of data-driven approaches, deriving its constants by fitting experimental data. Noteworthily, concerning machine learning, a rather small training dataset was found sufficient to obtain a qualified artificial neural network with very low validation mean squared error.

Further, it is worth underlying that the paper aimed to develop and compare various approaches to support the designer in the preliminary selection of strut-based lattice structures with the mechanical properties necessary for the application under investigation. Accordingly, it was preferred to keep all the approaches (including the constitutive model for FEA) as simple as possible to avoid high preparation and/or analysis and/or implementation times. However, future research should consider the development of more accurate FEA models for the description of the behavior of lattice structures manufactured by FDM 3D printing.

Finally, the possibility of extending the predictions to materials other than PLA should be considered. The work here presented is still limited to the case of PLA lattices, but the same strategies may be in principle applied to any other material. Similarly, 3D printing parameters such as printing speed, nozzle temperature or nozzle diameter are considered fixed in this work, but the data-driven approaches may be valuable also in view of exploring the effects of these variables on the mechanical properties of lattice structures.

In conclusion, the strategies presented in this paper enable the prediction of strength and stiffness of untested lattice configurations based on their geometrical features, which could be exploited early on during the design stage for faster examination of a large number of possible designs, but also later on to support design optimization. Depending on the designer's needs, the best approach can be selected based on the benefits and drawback highlighted throughout this work.

**Acknowledgments**

This work was funded by the European Union ERC CoDe4Bio Grant ID 101039467. Views and opinions expressed are however those of the author(s) only and do not necessarily reflect those of the European Union or the European Research Council. Neither the European Union nor the granting authority can be held responsible for them.

**Funding**

Open access funding provided by Università degli Studi di Pavia within the CRUI-CARE Agreement.

**Data Availability**

Data generated in this study have been deposited in the Zenodo database at https://doi.org/10.5281/zenodo.13853988.

**Publisher's Note** Springer Nature remains neutral with regard to jurisdictional claims in published maps and institutional affiliations.